\documentclass[a4paper, 12pt]{article}
\usepackage{amsmath}
\usepackage{amsfonts}
\usepackage{mathrsfs}
\usepackage{graphicx}
\usepackage{amsthm}
\usepackage{latexsym}
\usepackage{amssymb}
\newcommand{\lC}{\mathrm{l\hspace{-2.1mm}C}}
\newcommand{\lR}{\mathrm{I\hspace{-0.7mm}R}}

\numberwithin{equation}{section}

\usepackage{hyperref}

\begin{document}

%
%



\title{Deformation of Dyonic Black
Holes and Vacuum Geometries in Four Dimensional $N=1$
Supergravity}

\author{Bobby E. Gunara$^{\flat}$, Freddy P. Zen$^{\flat}$, and
Arianto$^{\flat, \sharp, \ddagger}$ \footnote{bobby@fi.itb.ac.id,
fpzen@fi.itb.ac.id}}

\date{}

\maketitle


\begin{center}
$^{\flat}$Indonesia Center for Theoretical and Mathematical Physics (ICTMP)\\
and \\
Theoretical Physics Lab., THEPI Devision,\\
Faculty of Mathematics and Natural Sciences,\\
 Institut Teknologi Bandung,\\
Jl. Ganesha 10 Bandung 40132, INDONESIA. \\
$^{\sharp}$Department of Physics, Udayana University\\
Jl. Kampus Bukit Jimbaran Kuta-Bali 80361, INDONESIA.\\
$^{\ddagger}$ Deceased
\end{center}


\begin{abstract}
We study some aspects of spherical symmetric dyonic
non-supersymmetric black holes in $4d \; N=1$ supergravity coupled
to chiral and vector multiplets on K\"ahler-Ricci solitons. Then,
we have a family of dyonic non-supersymmetric black holes deformed
with respect to the flow parameter related to the K\"ahler-Ricci
soliton, which possibly controls the nature of black holes, such
as their asymptotic  and near horizon geometries. Two types of
black holes are discussed, namely a family of dyonic
Reissner-Nordstr\"om-like black holes and Bertotti-Robinson-like
black holes where the scalars are freezing all over spacetime and
at the horizon, respectively. In addition, the corresponding
vacuum structures for such black holes are also studied in the
context of Morse-(Bott) theory. Finally, we give some simple
$\;{\lC \mathrm{P}}^n$-models whose superpotential and gauge
couplings have a linear form.
\end{abstract}

{\bf Keywords :}  Black Holes, Supergravity, Morse Theory,
K\"ahler-Ricci Flow




\section{Introduction and Conclusions}
\label{Intro}

This paper is devoted to extend our previous results in domain
walls \cite{GZ} to the case of black holes. In particular, the
black holes are non-supersymmetric, dyonic, and admit a spherical
symmetry. Such a class of solutions exists in $4d \; N=1$
supergravity coupled to chiral and vector multiplets where the
scalar manifold can be thought of as K\"ahler-Ricci (KR) soliton
\cite{cao} deformed with respect to $\tau \in \lR$. As we will
see, the setup further implies that the asymptotic geometry of
black holes may change as the flow parameter $\tau$ varies. Such
an object is called a family of black holes.\\
\indent The black holes are the solution of a set of equations of
motions such as the Einstein field equation, the gauge field and
the scalar field equations of motions which can be obtained by
varying the $N=1$ supergravity action with respect to the metric,
gauge fields, and scalar fields together with an additional
necessary condition, namely the variation of the flow parameter
$\tau$ vanishes. Moreover, the equations of motions show that
there is an additional potential called black hole potential
\cite{FGK} beside the scalar potential which play an important
role in analyzing the nature of black holes and the corresponding
vacuum structures on which the scalars become frozen in some
particular regions in spacetime, such as at the horizons, in
asymptotic geometries, or all over spacetime. Such black holes are
in general non-extreme. The word ``extreme" here is coming from
some studies of supersymmetric black holes in the context of $4d$
$N=2$ supergravity \cite{FGK} and $N=1$
supergravity\cite{ADFT1}.\\
\indent To be precise, we mainly focus on two classes of dyonic
spherical symmetric black holes. First, it is called a family of
dyonic Reissner-Nordstr\"om (RN)-like black holes on which the
scalar fields are frozen with respect to the spacetime coordinates
everywhere but they generally depend on the electric and magnetic
charges, and $\tau$. The asymptotic geometries are a family of
four dimensional Einsteinian geometries, namely de Sitter (dS),
Minkowski, or anti-de Sitter (AdS), which is $\tau$-dependent. In
other words, the KR soliton interpolates the dyonic RN-like black
holes with various asymptotic geometries. Here, the ground states
are generally the critical points of both the black
hole and the scalar potentials.\\
\indent The second kind is called a family of dyonic
Bertotti-Robinson (BR)-like black holes. Such black holes occurred
near the horizon, and their geometry is a product of two families
of surfaces, namely the families of two dimensional surfaces and
two-spheres. Here, the scalars are fixed at the horizon but they
generally depend on the electric and magnetic charges, and the
flow parameter $\tau$. The two dimensional surfaces are in general
the collection of Einsteinian surfaces. Imposing the positivity of
the entropy together with the time-like condition of a Killing
vector we then have only the AdS surfaces in the case at hand. In
addition, the vacuum structures can be viewed as the critical
points of the so called effective black
hole potential, formed by the black hole and the scalar potentials.\\
\indent The organization of this paper can be mentioned as
follows. In Section \ref{SUGRAKR} we discuss some aspects of four
dimensional $N=1$ supergravity coupled to chiral and vector
multiplets on KR solitons including the equations of motions. In
Section \ref{DBHV} we exploit some properties of dyonic RN-like
and black BR-like holes together with their vacua. Finally,  we
give some simple models for those types of black holes in Section
\ref{simple model}.

\section{$N=1$ Supergravity Coupled to Vector and Chiral
 Multiplets on the KR Soliton}
 \label{SUGRAKR}

In this section we consider some properties of four dimensional
$N=1$ supergravity coupled to vector and chiral multiplets where
the scalar manifold can be viewed as a one-parameter family of
K\"ahler manifolds generated by a KR soliton, satisfying
\cite{cao}
\begin{equation}
\frac{\partial g_{i\bar{j}}}{\partial \tau}(z, \bar{z}; \tau) = -2
R_{i\bar{j}}(z, \bar{z}; \tau),\quad 0 \le \tau <T \;,\label{KRF}
\end{equation}
for finite $T$ where the indices $i,j$ run from 1 to the dimension
of the KR soliton. As mentioned above, such a soliton can be
regarded as a volume deformation of K\"ahler manifolds and
further, affects the geometrical nature of a solution. For
example, as we will discuss in Section \ref{DBHV} and give an
evidence in Section \ref{simple model}, the use of KR solitons in
the context of black holes may change the black hole's asymptotic
geometries. Here, for the sake of simplicity we only consider the
``ungauged" case in the sense that we omit the couplings coming
from the isometries of the KR soliton. Additionally, we write
some terms which are useful for our analysis in the paper.\\
\indent The $N=1$ theory contains a gravitational multiplet
coupled with $n_v$ vector and $n_c$ chiral multiplets. An
interested reader can further read, for example, \cite{DF}. The
construction of the local $N=1$ theory on a KR soliton is as
follows. First, we consider the Lagrangian in \cite{DF} as the
initial Lagrangian at $\tau = 0$. Then, by replacing all couplings
that depend on the geometric quantities, such as the metric
$g_{i\bar{j}}(0)$ by the soliton $g^{i\bar{j}}(\tau)$, the bosonic
parts of the Lagrangian become
\begin{eqnarray}
{\mathcal{L}}(\tau) = -\frac{M_P^2}{2}R + {\mathcal{R}}_{\Lambda
\Sigma}\,{\mathcal{F}}_{\mu\nu}^{\Lambda}{\mathcal{F}}^{\Sigma|\mu\nu}
+ {\mathcal{I}}_{\Lambda \Sigma}\,
{\mathcal{F}}_{\mu\nu}^{\Lambda}\widetilde{\mathcal{F}}^{\Sigma|\mu\nu}
 + \, g_{i\bar{j}}(z, \bar{z}; \tau)\,
\partial_{\nu} z^i \,
\partial^{\nu}\bar{z}^{\bar{j}} - V(z,\bar{z}; \tau)\;, \label{L}
\end{eqnarray}
where $i,j = 1,...,n_c$, $M_P$ is Planck mass, and $R$ is the
Ricci scalar of the four dimensional spacetime. The scalar fields
$(z,\bar{z})$ span a Hodge-K\"ahler manifold endowed with the
metric $g_{i\bar{j}}(z, \bar{z}; \tau) \equiv
\partial_i
\partial_{\bar{j}}K(z,\bar{z}; \tau)$ satisfying (\ref{KRF}) with
 $K(z,\bar{z}; \tau)$ is a real function, called the K\"ahler
potential.  ${\mathcal{F}}^{\Lambda}_{\mu\nu}$ and
$\widetilde{\mathcal{F}}^{\Lambda}_{\mu\nu}$ are the gauge field
strength and its dual, respectively. While the gauge couplings
${\mathcal{R}}_{\Lambda \Sigma}$ and ${\mathcal{I}}_{\Lambda
\Sigma}$ are the real and the imaginary parts of the holomorphic
gauge couplings ${\mathcal{N}}_{\Lambda\Sigma} \equiv
{\mathcal{N}}_{\Lambda\Sigma}(z)$, with $\Lambda, \Sigma =
1,...,n_v$.  In order to have a consistent theory the gauge
coupling matrix ${\mathcal{N}}$ must be invertible. The $N=1$
scalar potential $V(z,\bar{z}; \tau)$ can be written down in terms
of a holomorphic superpotential $W$ \cite{GZ, DF}
\begin{equation}
 V(z,\bar{z}; \tau) = e^{K(\tau)/M^2_P}\left(g^{i\bar{j}}(\tau)\nabla_i W\,
 \bar{\nabla}_{\bar{j}} \bar{W}
 - \frac{3}{M^2_P} W \bar{W} \right)\;,
\label{V01}
\end{equation}
where $W$ is a holomorphic superpotential, $K(\tau) \equiv
K(z,\bar{z}; \tau)$, and $\nabla_i W\equiv
\partial_i W + (K_i(\tau)/M^2_P) W$.\\
\indent Let us first discuss how to derive in general a set of
field equations of motions obtained from the action denoted by
$S(\tau)$ related to the Lagrangian (\ref{L}). Similar as in the
standard procedure, we vary the action with respect to the bosonic
and fermionic fields.  Since the action also depends on the flow
parameter $\tau$, we also have to employ the variation of $\tau$.
As mentioned above, all fermions vanish at the level of the
equation of motions, thus we just focus on the variation of the
bosonic fields, namely $g_{\mu\nu}$, $A^{\Lambda}_{\mu}$, $z^i$,
$\bar{z}^{\bar{i}}$, and of the parameter $\tau$. Taking the
general condition $\delta S(\tau) =0$ and assuming $\delta \tau
=0$, the variation of $S(\tau)$ with respect to $g_{\mu\nu}$,
$A^{\Lambda}_{\mu}$, $z^i$, and $\bar{z}^{\bar{i}}$ should be
vanished in order to obtain the usual equations of motions, namely
the Einstein field equation
\begin{eqnarray}
 R_{\mu\nu} - \frac{1}{2} g_{\mu\nu} R &=& g_{i\bar{j}}(\tau)
 (\partial_{\mu}z^i \partial_{\nu}\bar{z}^{\bar{j}}
 + \partial_{\nu}z^i \partial_{\mu}\bar{z}^{\bar{j}}) - g_{i\bar{j}}(\tau) \, g_{\mu\nu}
\partial_{\rho}z^i \partial^{\rho}\bar{z}^{\bar{j}} \nonumber\\
&& + \, 4 \, {\mathcal{R}}_{\Lambda\Sigma}
\,{\mathcal{F}}_{\mu\rho}^{\Lambda}{\mathcal{F}}^{\Sigma}_{\nu\sigma}
g^{\rho\sigma} - g_{\mu\nu} {\mathcal{R}}_{\Lambda\Sigma}
\,{\mathcal{F}}_{\rho\sigma}^{\Lambda}{\mathcal{F}}^{\Sigma|\rho\sigma}
 + g_{\mu\nu} V(\tau) \;, \label{Einsteineq}
\end{eqnarray}
the Maxwell equation
\begin{equation}
 \partial_{\nu} \left(\varepsilon^{\mu\nu\rho\sigma} \sqrt{-g}
 \, {\mathcal{G}}_{\Lambda|\rho\sigma}\right) = 0\;,
\label{gaugeEOM}
\end{equation}
where
\begin{equation}
{\mathcal{G}}_{\Lambda|\rho\sigma}  \equiv {\mathcal{I}}_{\Lambda
\Sigma} {\mathcal{F}}^{\Sigma|\rho\sigma} - {\mathcal{R}}_{\Lambda
\Sigma} \widetilde{\mathcal{F}}^{\Sigma}_{\rho\sigma} \;,
\label{magnetfield}
\end{equation}
and the scalar field equation of motions
\begin{equation}
 \frac{g_{i\bar{j}}(\tau)}{\sqrt{-g}} \, \partial_{\mu} \left( \sqrt{-g} \,g^{\mu\nu}
 \partial_{\nu}\bar{z}^{\bar{j}}
 \right) + \bar{\partial}_{\bar{k}} g_{i\bar{j}}(\tau) \, \partial_{\nu}\bar{z}^{\bar{j}}
 \partial^{\nu}\bar{z}^{\bar{k}} = \partial_i {\mathcal{R}}_{\Lambda
\Sigma}\,{\mathcal{F}}_{\mu\nu}^{\Lambda}{\mathcal{F}}^{\Sigma|\mu\nu}
+ \partial_i{\mathcal{I}}_{\Lambda \Sigma}\,
{\mathcal{F}}_{\mu\nu}^{\Lambda}\widetilde{\mathcal{F}}^{\Sigma|\mu\nu}
- \partial_i V(\tau)\;, \label{scalarEOM}
\end{equation}
respectively. \\
%
\indent Our interest is then to solve the equations of motions
mentioned above on the background where the four dimensional
metric ansatz is static and spherical symmetric,
  given by
\begin{equation}
 ds^2 = e^{A(r, \tau)}\, dt^2 - e^{B(r, \tau)}\, dr^2 - e^{C(r, \tau)}\,
 (d\theta^2 + {\mathrm{sin}}^2\theta \, d\phi^2)\;,
\label{metricans}
\end{equation}
with $A(r, \tau), B(r, \tau)$, and $C(r, \tau)$ are generally
assumed to be $\tau$-dependent functions. In other words, the
above ansatz defines a one-parameter family of spherical static
metrics with respect to $\tau$. It is worth mentioning that in
general a redefinition of the radial coordinate $r$ cannot be
employed since all of the functions vary with respect to $\tau$
and in addition, the ansatz (\ref{metricans}) may have a
singularity at finite $\tau$. However, there are two particular
situations where one can redefine $r$.  The first case is when all
functions do not depend on $\tau$, while the second case is when
$r$ is redefined at a particular value of $\tau$. The later case
is because we have a one-parameter family of spherical symmetric
geometries in which there possibly exists a topological changes of
the shape of the geometries described by the ansatz
(\ref{metricans}) with respect to $\tau$. We will come back to
this issue in Section \ref{RNBH}.\\
\indent Let us investigate the  field equations of motions
(\ref{Einsteineq})-(\ref{scalarEOM}). In the gauge field sector we
simply
 take a case where the nonzero
electromagnetic field strength components are
${\mathcal{F}}^{\Lambda}_{01}$ and ${\mathcal{F}}^{\Lambda}_{23}$.
Solving the gauge field Bianchi identities, we find
\begin{eqnarray}
{\mathcal{F}}^{\Lambda}_{01} &=& \frac{1}{2} \,
e^{\frac{1}{2}(A+B)-C}\,
({\mathcal{R}}^{-1})^{\Lambda\Sigma}({\mathcal{I}}_{\Sigma\Gamma}\,
g^{\Gamma} - q_{\Sigma}) \;,\nonumber\\
{\mathcal{F}}^{\Lambda}_{23} &=& - \frac{1}{2} \, g^{\Lambda} \,
{\mathrm{sin}}\theta  \;, \label{solgaugeEOM}
\end{eqnarray}
Then, the field equations of motions are reduced into three
independent equations
\begin{eqnarray}
- \frac{1}{2}\, C'\, \left(\frac{1}{2}\, C' + A'\right) + e^{B-C}
= - g_{i\bar{j}}(\tau) z^i{'} \bar{z}^{\bar{j}}{'} + e^B
\left(e^{-2C} V_{\mathrm{BH}} + V (\tau) \right) \;,
\nonumber\\
 - \frac{1}{2}\,e^{-B}\left(A'' + C'' + \frac{1}{2} (A'+C')(A'-B')
 + \frac{1}{2} \, {C'}^2 \right)
 = e^{-B} \,g_{i\bar{j}}(\tau) z^i{'} \bar{z}^{\bar{j}}{'}
  - e^{-2C} V_{\mathrm{BH}} + V (\tau)  \;, \nonumber\\
 g_{i\bar{j}}(\tau) \, \bar{z}^{\bar{j}}{''} + \bar{\partial}_{\bar{k}}
 g_{i\bar{j}}(\tau) \, \bar{z}^{\bar{j}}{'} \bar{z}^{\bar{k}}{'}
 + \frac{1}{2} \left(A' - B' + 2C' \right)g_{i\bar{j}}(\tau) \,
 \bar{z}^{\bar{j}}{'} = e^{B} \left( e^{-2C} \partial_i V_{\mathrm{BH}}
 + \partial_i V (\tau)\right) \;, \nonumber\\
\label{fieldEOM}
\end{eqnarray}
where $z^i = z^i(r, \tau)$ and $z^i{'} \equiv
\partial z^i / \partial r$. The function $V(\tau)$ is the scalar potential
in (\ref{L}), whereas the scalar function $V_{\mathrm{BH}}$ is
called the black hole potential \cite{FGK}
\begin{eqnarray}
V_{\mathrm{BH}} \equiv - \frac{1}{2} \, (g \:\, q ) \;
\left(\begin{array}{cc} {\mathcal{R}} + {\mathcal{I}}
\,{\mathcal{R}}^{-1} \,{\mathcal{I}} & -
{\mathcal{I}}\, {\mathcal{R}}^{-1} \\
- {\mathcal{R}}^{-1}\, {\mathcal{I}} & {\mathcal{R}}^{-1}
\end{array} \right)  \; \left(\begin{array}{c}  g \\
 q \end{array} \right)
 \;.
 \label{VBH}
\end{eqnarray}
 Note that in this case we have $V_{\mathrm{BH}} \ge 0$, and $V(\tau)
\in {\lR}$ is assumed to be well-defined for finite $\tau \ne
\tau_0$, where $\tau_0$ is a singular point of the geometric flow
in (\ref{KRF}). In this paper we assume that $V_{\mathrm{BH}}= 0$
if all charges vanish.

\section{Dyonic Black Holes and Their Vacuum Structure}
\label{DBHV}

\subsection{Dyonic RN-like Black Holes}
\label{RNBH}

Within the ansatz (\ref{metricans}), in order to obtain dyonic
RN-like black holes one has to set
\begin{eqnarray}
z^i{'} = 0 \;, \quad
\partial_i V_{\mathrm{BH}} = 0 \;, \quad
 \partial_i V (\tau) = 0 \;,\label{extremcon}
\end{eqnarray}
all over spacetime. The first and the second equations in
(\ref{extremcon}) have been considered previously in \cite{ADFT1}
without introducing KR soliton and the scalar potential $V \equiv
0$. The second and the third equations in (\ref{extremcon})
defines critical points of the potentials
 in which we have generally $\tilde{z}^i_0(g,
q; \tau)$.\\
\indent Additionally, in this region the function $C(r, \tau)$
simply takes the form
\begin{equation}
C(r, \tau) = 2 \, {\mathrm{ln}} r + {\mathrm{ln}}
\hat{\sigma}(\tau)
 \;, \label{Cf}
\end{equation}
where $\hat{\sigma}(\tau)$ is an arbitrary function of the flow
parameter $\tau$, related to the area deformation of the
two-sphere $S^2$. In other words, $2n_c$-dimensional KR solitons
described by (\ref{KRF}) may induce an area deformation of $S^2$
which might also cause a topological change of the black hole
geometry. However, the form of $C(r, \tau)$ in (\ref{Cf}) shows
that we always have $\hat{\sigma}(\tau) > 0$ for  all $\tau$. Note
that if $\hat{\sigma}(\tau) =1$, then the black hole geometry is
indeed RN with frozen scalar fields.\\
\indent Now, we discuss the solution of the equation of motions
(\ref{fieldEOM}) in the extremal condition (\ref{extremcon}).
Inserting (\ref{extremcon}) and (\ref{Cf}) into (\ref{fieldEOM})
and after some manipulations we obtain

\newtheorem{metricsolRN}{Theorem}[]
\begin{metricsolRN}\label{metricsolRN}
Suppose $p_0 \equiv \big(z_0(g,q;\tau), \bar{z}_0(g,q;\tau)\big)$
is a vacuum satisfying (\ref{extremcon}). Then, at $p_0$ the
solution of the equation of motions (\ref{fieldEOM}) has the form
\begin{equation}
 ds^2 = \Delta(r, \tau)\, dt^2 - \Delta(r, \tau)^{-1}\, dr^2 - \hat{\sigma}(\tau) r^2\,
 (d\theta^2 + {\mathrm{sin}}^2\theta \, d\phi^2)\;,
\label{metricsol}
\end{equation}
with
\begin{equation}
\Delta(r, \tau) = \hat{\sigma}(\tau)^{-1} -\frac{2M(\tau)}{r} +
\frac{V^0_{\mathrm{BH}}(\tau)}{\hat{\sigma}^2(\tau)\, r^2} -
\frac{1}{3} V_0(\tau)\, r^2 \;, \label{Ar}
\end{equation}
where we have defined $V^0_{\mathrm{BH}}(\tau) \equiv
V_{\mathrm{BH}}(p_0)$ and $ V_0(\tau) \equiv  V(p_0; \tau)$.
\end{metricsolRN}
Note that the metric (\ref{metricsol}) does not preserve
supersymmetry. $M(\tau)$ is the black hole's mass and $M(\tau) \ge
0$. Meanwhile the black hole potential $V^0_{\mathrm{BH}}(\tau) >
0$ for all $\tau$ and $V^0_{\mathrm{BH}}(\tau)  = 0$ when all
charges vanish. In the latter case, the black hole geometry is a
one-parameter family of
four dimensional Einstein manifolds.\\
\indent Some comments are as follows. In general the cosmological
constant $V_0(\tau)$ may deform with respect to $\tau$ for $\tau
\ne \tau_0$. We have three different backgrounds, namely dS for
$V_0(\tau) >0$, Minkowskian for $V_0(\tau) = 0$, and AdS for
$V_0(\tau) < 0$. In addition, the metric (\ref{metricsol})
generally becomes Einsteinian in asymptotic region, namely around
$r \to +\infty$, for finite
$M(\tau)$, $V^0_{\mathrm{BH}}(\tau)$ and $V_0(\tau)$.\\
\indent Let us now discuss the ground states related to the
RN-like metric (\ref{metricsol}). First, we introduce some
definitions as follows.

\newtheorem{MV}[metricsolRN]{Definition}
\begin{MV}\label{MV}
A subset ${\bf{M}}_v$ that describes a collection of critical
points of the scalar potential $V(\tau)$, is defined as
\begin{equation}
{\bf{M}}_v \equiv \big\{\hat{p}_0 \in {\bf{M}} \, | \, \partial_i
V(\hat{p}_0) =0 \big\} \subset {\bf{M}} \;, \label{MV0}
\end{equation}
where $\,{\bf{M}}$ is a K\"ahler manifold generated by (\ref{KRF})
and $\hat{p}_0 \equiv \big(\hat{z}_0(\tau),
\bar{\hat{z}}_0(\tau)\big)$ characterized by the $\tau$-dependent
the dimension and the index $(m_v(\tau), \lambda_v(\tau))$,
 respectively.
\end{MV}

\newtheorem{MVBH}[metricsolRN]{Definition}
\begin{MVBH}\label{MVBH}
A subset ${\bf{M}}_b$ that represents a collection of critical
points of the black hole potential $V_{\mathrm{BH}}$, is defined
as
\begin{equation}
{\bf{M}}_b \equiv \big\{\tilde{p}_0 \in {\bf{M}} \, | \,
\partial_i V_{\mathrm{BH}}(\tilde{p}_0)
 =0 \big\} \subset {\bf{M}} \;, \label{MVBH0}
\end{equation}
 where $\tilde{p}_0 \equiv \big(\tilde{z}_0(g,q), \bar{\tilde{z}}_0(g,q) \big)$.
 Each $\tilde{p}_0$ of ${\bf{M}}_b$ has the quantities $(m_b, \lambda_b)$ where $m_b \equiv m_b(g,q) $
 and $\lambda_b \equiv \lambda_b(g,q)$ are the dimension and the index of $\,{\bf{M}}_b$,
 respectively.
\end{MVBH}
\indent Thus, from the above statements we can write down the
following results.

\newtheorem{MVac}[metricsolRN]{Theorem}
\begin{MVac}\label{MVac}
Suppose both $\, V(\tau)$ and $V_{\mathrm{BH}}$ are nonconstant.
Then, there exists some $p_0 \equiv \big(z_0(g,q;\tau),
\bar{z}_0(g,q;\tau)\big)$ which belong to
\begin{equation}
{\bf{M}}_{\circ} \equiv \big\{p_0 \in {\bf{M}} \, | \,
\partial_i V(p_0) = \partial_i V_{\mathrm{BH}}
(p_0) =0 \big\} \simeq {\bf{M}}_b \cap {\bf{M}}_v \;,
\label{Mvacdef}
\end{equation}
describing (\ref{extremcon}) whose dimension and index of each
$p_0$ are respectively given by $(m_{\circ}, \lambda_{\circ})$
with $m_{\circ}(g,q;\tau)$ and $\lambda_{\circ}(g,q;\tau)$.
\end{MVac}
Some comments are as follows. First, in general we have a
situation where $z_0(g,q;\tau)$ which is referred to as a dynamic
case. Such a situation occurs if
 the scalar potential $V(\tau)$ is at least nonconstant. If $V(\tau)
 =0$ for all $z$ and $\tau$, then $(m_{\circ}, \lambda_{\circ})$ are independent of
 $\tau$ which is referred to as a static case. Second, if
 ${\bf{M}}_{\circ}$ is an empty set, then we may possibly have a case
 that will be discussed in Subsection \ref{dyonBMFY}.

\subsection{Dyonic BR-like Black Holes}
 \label{dyonBMFY}

Near the horizon we have $z^i{'} =0$, and the metric
(\ref{metricans}) in this limit becomes a product of two families
of surfaces, namely $M^{1,1} \times S^2$, where $M^{1,1}$ and
$S^2$ are families of two dimensional surfaces and two-spheres,
respectively. Moreover, the time-like condition of the Killing
vector $\xi = \frac{\partial}{\partial t}$ implies $e^{A(r, \tau)}
> 0$.\\
 \indent In the near-horizon limit,
equations (\ref{fieldEOM}) read
\begin{eqnarray}
\frac{1}{r_h^2} = \frac{1}{r_h^4} V^h_{\mathrm{BH}} + V_h(\tau)
\;, \quad \ell = \frac{1}{r_h^4} V^h_{\mathrm{BH}} - V_h(\tau) \;,
\quad
 \left(\frac{1}{r_h^4} \frac{\partial V_{\mathrm{BH}}}{\partial
z^i} + \frac{\partial V(\tau)}{\partial z^i} \right)(p_h) = 0
 \;,\label{fieldEOMhor}
\end{eqnarray}
with $\ell \equiv \ell(g, q; \tau)$ and $p_h \equiv (z_h,
\bar{z}_h)$. We have defined $V^h_{\mathrm{BH}} \equiv
V_{\mathrm{BH}} (p_h)$, $V_h(\tau) \equiv  V(p_h; \tau)$, and
$\lim_{r \to r_h} z^i \equiv z^i_h$. The solutions of
(\ref{fieldEOMhor}) are given by
\begin{eqnarray}
r_h^2 =  V^h_{\mathrm{eff}}(\tau) \;, \quad
 \ell^{-1} =
\frac{V^h_{\mathrm{eff}}(\tau)}{\sqrt{1 -4 V^h_{\mathrm{BH}}
V_h(\tau)}}
 \;, \quad
\frac{\partial V_{\mathrm{eff}}}{\partial z^i} (p_h)  = 0 \;,
\label{solfieldEOMhor}
\end{eqnarray}
where
\begin{equation}
 V_{\mathrm{eff}}(\tau) \equiv  \frac{1-\sqrt{1-4 V_{\mathrm{BH}}V(\tau)}}{2
 V(\tau)}\;\label{effV}
\end{equation}
is the effective black hole potential \cite{BFMY} and
$V^h_{\mathrm{eff}}(\tau) \equiv V_{\mathrm{eff}}(p_h; \tau)$. The
last equation in (\ref{solfieldEOMhor}) shows that near the
horizon the scalars $z_h$ can be viewed as the critical points of
$V_{\mathrm{eff}}$ in the scalar manifold ${\bf{M}}$ and moreover,
we have $z_h \equiv z_h(g,q; \tau)$. In this case the black hole
entropy depends linearly on $V^h_{\mathrm{eff}}(\tau)$
\cite{BFMY}. \\
\indent Using the above results, we can write the following
statement:

\newtheorem{ADS2}[metricsolRN]{Theorem}
\begin{ADS2}\label{ADS2}
For $B= \pm A$, the entropy of BR black holes is strictly positive
with $e^{A(r, \tau)} > 0$ if and only if $M^{1,1}$ is $AdS_{1,1}$.
\end{ADS2}
\indent Let us make some remarks here. As observed in \cite{BFMY},
in the case at hand the spacetime is not conformally flat since
the radius of $AdS_{1,1}$ given by $r_a \equiv \ell^{-1/2}$ does
not equal to the radius of $S^2$, namely $r_h$. The results in
Theorem \ref{ADS2} satisfy the general near horizon condition
without introducing the K\"ahler-Ricci flow \cite{KLR} excluding
the singular flat case. Finally, the positivity of the entropy of
the model restricts $r^2_h > 0$ for all $\tau$ and $\tau \ne
\tau_0$ which means that the KR flow keeps the topology of $S^2$
for all $\tau$ and $\tau \ne \tau_0$.\\
\indent Now, we turn to discuss the definition of an effective
vacuum geometries as follows.

\newtheorem{MVeff}[metricsolRN]{Definition}
\begin{MVeff}\label{MVeff}
A subset ${\bf{M}}_e$ defined as
\begin{equation}
{\bf{M}}_e \equiv \big\{p_h \in {\bf{M}} \, | \,
\partial_i V_{\mathrm{eff}}(p_h) = 0 \;; V^h_{\mathrm{eff}} > 0 \big\}
\subset {\bf{M}} \;, \label{MVeff0}
\end{equation}
is called the effective vacuum structure.
\end{MVeff}
\noindent It is important to notice that in general the points
$p_h \ne \hat{p}_0$ where $\hat{p}_0$ is given by (\ref{Mvacdef})
and thus, $p_h$ are not the critical points of the scalar
potential $V(\tau)$ and the black
hole potential $V_{\mathrm{BH}}$.\\
\indent We have then

\newtheorem{MVac3}[metricsolRN]{Theorem}
\begin{MVac3}\label{MVac3}
If both $V(\tau)$ and $V_{\mathrm{BH}}$ are nonconstant, then each
$p_h$ of $\,{\bf{M}}_e$ is characterized by the dimension
$m_e(g,q; \tau)$ and the index $\lambda(g,q; \tau)$.
\end{MVac3}
Some comments are as follows. $p_h$ is said to be nondegenerate or
isolated, if $H_{V_{\mathrm{eff}}}$ admits only non zero
eigenvalues which follows from $m_e = 0$ and $\lambda_e$ is a
Morse index. In particular, if all eigenvalues of
$H_{V_{\mathrm{eff}}}$ are strictly positive, then $p_h$ is called
an attractor \cite{BFMY}. Finally, for $V(\tau) \to 0$, so we have
$\,{\bf{M}}_e \simeq {\bf{M}}_b$ where $\, {\bf{M}}_b$ is
described in the Definition \ref{MVBH}.

\section{Simple $\; {\mathrm{l\hspace{-2.8mm}C}}\mathrm{P}^n$-Models}
\label{simple model}

In this section we consider a simple $\;{\lC \mathrm{P}}^n$-model
whose gauge couplings and superpotential have a linear form,
namely
\begin{eqnarray}
{\mathcal{N}}_{\Lambda\Sigma}(z) = (b_0 + b_i z^i)
\delta_{\Lambda\Sigma} \;, \quad
 W(z) = a_0 + a_i z^i \;, \label{gcoupw}
\end{eqnarray}
respectively, with $a_0, a_i, b_0, b_i \in \lR$. The K\"ahler
potential in this case is given by
\begin{equation}
 K(\tau) = \sigma(\tau) \, {\mathrm{ln}}(1 + \vert z \vert^2) \;, \label{KpotCPn}
\end{equation}
where $\sigma(\tau) \equiv 1- 2 (n_c +1) \tau$ and $\vert z
\vert^2 \equiv \delta_{i\bar{j}} \, z^i \bar{z}^{\bar{j}}$. In
addition,
we simply set $\hat{\sigma}(\tau) = 1$ for all $\tau$.\\
\indent Let us first begin to construct the case for RN-like black
holes where at the vacua we have  $b_i = 0$ for all $z$ coming
from the first derivative of the gauge couplings, and $\partial_i
V (p_0; \tau) = 0$. For the case at hand, the black hole potential
$V_{\mathrm{BH}}$ at $p_0$ is positive with $b_0 >0$ which follows
that the vacuum
geometry is  ${\bf{M}}_v \subset {\bf{M}}$ .\\
\indent First, we simply take $z_0 = 0$. This case can be split
into two cases as follows. The first case is $a_i = 0$ for all $i$
and $a_0 \ne 0$. In this case the scalar potential (\ref{V01}) is
then negative and independent of $\tau$ whose Hessian matrix is
simply a $\tau$-dependent diagonal matrix. Thus we have an AdS
black hole with isolated ground states whose Morse index is $2n_c$
for $\tau < 1/2(n_c + 1)$, and they change to $0$ for $\tau
> 1/2(n_c + 1)$, as the geometry evolves with respect to the
KR equation (\ref{KRF}). On the other hand, the ground states turn
out to be degenerate Minkowskian spacetime
if $a_0 =0$.\\
\indent The second case is $a_i \ne 0$ for some $i$ and $a_0 = 0$.
The scalar potential (\ref{V01}) has the form
\begin{equation}
 V (0; \tau) = \sigma(\tau)^{-1} \, a^2  \;,
\end{equation}
where $a^2 \equiv \delta^{ij}a_i a_j$. The non-zero components of
the Hessian matrix of $V(\tau)$ is given by
\begin{equation}
\partial_i \bar{\partial}_{\bar{j}}V(0; \tau) = \left( 1
+ \frac{\sigma(\tau)}{M^2_P}\right) \frac{a^2}{\sigma(\tau)} \,
\delta_{ij} - \frac{a_i \, a_j}{M^2_P}   \;. \label{HessV1}
\end{equation}
In this case for $a^2 \ne 0$ we have dyonic black holes with three
different backgrounds, namely dS backgrounds for $\tau < 1/2(n_c +
1)$, AdS backgrounds for $\tau > 1/2(n_c + 1)$, and Minkowski
backgrounds around $\tau \to \pm \infty$. Note that the latter
case happens also for $a^2 = 0$. These situations are strong
evidences of our previous statements that the parameter
$\tau$ indeed controls the shape of the black hole geometry.\\
%
%
\indent Next, we consider a simple case for BR-like black holes
where $n_c =1$, $W(z) = a_0 \ne 0$, and $b_0 \gg b_1 \gtrsim 0$.
We set $gq =0$, and by simply taking that the effective vacuum
exists near the origin, namely $z_h \approx 0$, we obtain
\begin{eqnarray}
 x_h \approx  \frac{3a_0^2 \, b_1 \, g^2}{4 \, \sigma(\tau)} \, \left( b_0 \, g^2
 +\frac{q^2}{b_0 }
 \right)^{-1}
 \left(  \frac{a_0^2}{M_P^2} - 1 \right)^{-1}  \;, \quad
y_h = 0 \;, \label{critpointBR}
\end{eqnarray}
for large $M_P$ and $\tau$. The effective black hole potential
(\ref{effV}) has the form
\begin{eqnarray}
V_{\mathrm{eff}}(\tau) \approx \frac{M_P^2}{6 a_0^2} \,
\left\lbrack \sqrt{1 + \frac{12 a_0^2}{b_0 M_P^2} \Big(b_0^2 \,
g^2 + q^2 \Big)} - 1 \right\rbrack \;,\label{effV1}
\end{eqnarray}
which is already positive. Furthermore, the analysis on the
eigenvalues of $H_{V_{\mathrm{eff}}}$ shows that the vacua
(\ref{critpointBR}) are not attractors since all eigenvalues of
$H_{V_{\mathrm{eff}}}$ is strictly negative.

\vskip 1truecm

\hspace{-0.2 cm}{\Large \bf Acknowledgement}
\\
\vskip 0.15truecm \hspace{-0.6 cm} \noindent It is a pleasure to
acknowledge H. Alatas and T. Kimura for useful discussions. We
also greatly thanks M. Satriawan for proof reading the manuscript
and correcting some grammar. This work is supported by Riset KK
ITB 2009 No. 243/K01.7/PL/2009, ITB Alumni Association Research
Grant (HR IA-ITB) 2009 No. 180a/K01.7/PL/2009, ITB Alumni
Association Research Grant (HR IA-ITB) 2010 No.
1443b/K01.7/PL/2010, and ITB Alumni Association Research Grant (HR
IA-ITB) 2011 No. 2208c/I1.C01/PL/2011.

\end{document}